# A Kinetic Study of the Gas-Phase $O(^1D)$ + $CH_3OH$ and $O(^1D)$ + $CH_3CN$ Reactions. Low Temperature Rate Constants and Atomic Hydrogen Product Yields.

Kevin M. Hickson[1,*] and Jean-Christophe Loison[1]

[1]Univ. Bordeaux, CNRS, Bordeaux INP, ISM, UMR 5255, F-33400 Talence, France

**Abstract**

Atomic oxygen in its first excited singlet state, $O(^1D)$, is an important species in the photochemistry of several planetary atmospheres and has been predicted to be a potentially important reactive species on interstellar ices. Here, we report the results of a kinetic study of the reactions of $O(^1D)$ with methanol, $CH_3OH$, and acetonitrile, $CH_3CN$, over the 50-296 K temperature range. A continuous supersonic flow reactor was used to attain these low temperatures coupled with pulsed laser photolysis and pulsed laser induced fluorescence to generate and monitor $O(^1D)$ atoms respectively. Secondary experiments examining the atomic hydrogen product channels of these reactions were also performed, through laser induced fluorescence measurements of $H(^2S)$ atom formation. On the kinetics side, the rate constants for these reactions were seen to be large (> 2 × $10^{-10}$ $cm^3$ $s^{-1}$) and consistent with barrierless reactions, although they display contrasting dependences as a function of temperature. On the product formation side, both reactions are seen to yield non-negligible quantities of atomic hydrogen. For the $O(^1D)$ + $CH_3OH$ reaction, the derived yields are in good agreement with the conclusions of previous experimental and theoretical work. For the $O(^1D)$ + $CH_3CN$ reaction, whose H-atom formation channels had not previously been investigated, electronic structure calculations of several new product formation channels were performed to explain the observed H-atom yields. These calculations demonstrate the barrierless and exothermic nature of the relevant exit channels, confirming that atomic hydrogen is also an important product of the $O(^1D)$ + $CH_3CN$ reaction.

# 1 Introduction

The reactions of electronically excited atomic oxygen, O($^1$D), have been extensively studied over the years due to their importance in Earth's atmospheric chemistry. In particular, ozone photolysis at UV wavelengths in the stratosphere produces high yields of O($^1$D) atoms that react rapidly with trace gases producing reactive radicals such as OH and H which play important roles in catalytic ozone destruction cycles in the stratosphere. Methanol, $CH_3OH$, is the second most abundant organic molecule present in the atmosphere after $CH_4$ and is thought to be emitted mostly during the growth and decay of plants in addition to anthropogenic sources such as solvent use, automobile emissions and industrial processes.[1] Acetonitrile, $CH_3CN$, is also present in the atmosphere as a result of biomass burning and automobile emissions. In contrast to $CH_3OH$, which has a short atmospheric lifetime of only a few days,[1] $CH_3CN$ has a relatively long lifetime of more than a year due to its low inherent reactivity,[2] allowing trace amounts of $CH_3CN$ to be transported to the stratosphere. Both $CH_3OH$[3] and $CH_3CN$[4,5] have been positively identified in the stratosphere, where they can potentially react with O($^1$D) atoms.

The reactions of O($^1$D) are also potentially important in the chemistry of planetary atmospheres. O($^1$D) atoms are known to play a role in the chemistry of the Martian atmosphere through the photodissociation of $CO_2$ at VUV wavelengths which results in high yields of O($^1$D) atoms.[6] Here, the reactions of O($^1$D) with $H_2$ and $H_2O$ are important sources of OH radicals which regenerate $CO_2$ by reaction with CO. Although O($^1$D) reactions are unimportant in the gas-phase chemistry of the interstellar medium (ISM), there is significant interest in O($^1$D) atom reactions as potential sources of complex organic molecules (COMs) on interstellar ices. Recent experimental studies of the reactions of O($^1$D) with $CH_4$,[7] $C_2H_6$, $C_2H_4$ and $C_2H_2$[8] on interstellar ice analogues, have shown that these reactions can efficiently produce COMs such as $CH_3OH$, $C_2H_5OH$, $CH_3CHO$ and $CH_2(O)CH_2$. Astrochemical models are now beginning to incorporate such processes,[9,10] where O($^1$D) atoms are generated by the photodissociation/radiolysis of oxygen bearing molecules already present in the ices. As $CH_3OH$ typically accounts for 5 % of the ice composition (with respect to the major constituent $H_2O$)[11] the reaction between O($^1$D) and $CH_3OH$ could represent an important source of COMs in interstellar chemistry. Although $CH_3CN$ has only been detected in the gas-phase ISM so far,[12,13,14] it is considered to be present on icy interstellar dust grains where it could contribute to the formation of nitrogen bearing COMs.[15,16] Although we investigate these processes in the gas-phase and not on ice surfaces, gas-phase studies provide insights into their surface reactivity (if the rate constants are slow, this could indicate the presence of a barrier to product formation) and the associated theoretical calculations and experimental product studies elucidate the intermediate species formed which are the likely primary products in the equivalent surface reaction.

Mechanistically at room temperature and below, O($^1$D) atoms react with organic molecules predominantly by insertion into C-H and O-H bonds, forming long-lived intermediates such as $CH_3OH$ during the O($^1$D) + $CH_4$ reaction and $H_2O_2$ during the O($^1$D) + $H_2O$ reaction before dissociating.[17,18,19,20,21] When $CH_3OH$ is employed as the coreagent species, as a bifunctional molecule two possibilities present themselves for the insertion process. Here, O($^1$D) can insert into either one of the C-H bonds of the $CH_3$ moiety or into the O-H bond itself, leading to the possible formation of two distinct intermediate species, $HOCH_2OH$ following C-H insertion or $CH_3OOH$ formation following O-H insertion.[22,23,24] While OH radicals are observed as the major product species, studies performed by selectively substituting H-atoms for D-atoms[22,23,24] clearly indicate that OH radicals are preferentially formed by dissociation of the $CH_3OOH$ intermediate[22,24] whereas H-atoms are preferentially formed from the $HOCH_2OH$ intermediate (although this intermediate also contributes to OH formation), given that the $CH_3OOH$ dissociation towards both $CH_2OOH$ + H and $CH_3OO$ + H is an essentially thermoneutral process.[23] When $CH_3CN$ is used as the coreagent species, calculations indicate that reaction with O($^1$D) can occur by three different pathways; O atom insertion into C-H or C-C bonds and addition of atomic oxygen to the carbon atom of the CN group.[25]

     To evaluate the importance of the O($^1$D) + $CH_3OH$ and O($^1$D) + $CH_3CN$ reactions in atmospheric and interstellar chemistry, an experimental kinetic study has been performed at room temperature and below using a supersonic flow reactor coupled with pulsed laser photolysis and laser-induced fluorescence for the production and detection of O($^1$D) atoms. In addition, measurements of the temperature dependent H-atom product yields for the O($^1$D) + $CH_3OH$ reaction and the H-atom yield for the O($^1$D) + $CH_3CN$ reaction at room temperature have also been performed to elucidate the preferred product channels of these reactions. In addition to the experimental work, electronic structure calculations of the exit channels of the O($^1$D) + $CH_3CN$ reaction have also been undertaken. These calculations are complementary to those of a previous theoretical study of this reaction,[25] and have been performed to allow us to explain the experimental findings.

The paper is structured as follows: Section 2 describes the experimental methodology used to study these reactions, while Section 3 presents the experimental results and discusses them in the context of earlier work where available. Section 3 also provides a brief description of the supplementary theoretical calculations. The conclusions are presented in Section 4.

**2 Experimental Methods**

The experiments reported here were made using a supersonic flow type reactor (also known as CRESU – cinétique de réaction en écoulement supersonique uniforme or reaction kinetics in a uniform supersonic flow). A detailed description of this apparatus was provided in earlier publications.[26,27] The main feature of this system, the Laval nozzle, allows supersonic flows of a specified carrier gas to be generated in the reactor under vacuum. These flows, which are created by isentropic expansion through the nozzle possess uniform density, velocity and temperature profiles, providing suitable environments to study the kinetics of fast reactions at low temperature. The original apparatus as described has been modified over the years, to incorporate a tunable narrow-band VUV source for detection purposes. In this way, it has been possible to study the low temperature reactivity of several atomic radicals such as $C(^3P)$,[28,29] $O(^1D)$,[30,21] $N(^2D)$[31,32] and $C(^1D)$ (indirectly through product $H(^2S)$ formation)[33,34] by exciting electronic transitions of these species in the 115-130 nm wavelength range. Despite the availability of Laval nozzles based on several different carrier gases including $N_2$, $SF_6$ and Ar, the present experiments were restricted to using argon type nozzles as the rate constants for $O(^1D)$ quenching by Ar are small enough (6-7 × $10^{-13}$ cm$^3$ s$^{-1}$) to allow kinetics measurements to be performed.[30,35] Indeed, $N_2$ is known to rapidly quench $O(^1D)$ atoms at room temperature and below.[30,36] Three different nozzles were employed here, providing supersonic flows with characteristic temperatures of 127 K, 75 K and 50 K with Ar as the carrier gas. Other relevant properties for flows generated by these Laval nozzles are reported in Table 1 of Nuñez-Reyes and Hickson.[37] To increase the range of the measurements to higher temperatures, the Laval nozzle was removed from the system, thereby allowing experiments to be performed at room temperature (296 K).

As both of the coreagent molecules used here are liquids at room temperature, it was necessary to employ a reliable method to introduce a known quantity of their vapour into the supersonic flow. During these experiments, methanol vapour was added to the gas flow upstream of the Laval nozzle via a controlled evaporation mixing system (CEM). Here, liquid methanol was contained in a 1 litre stainless steel reservoir maintained at a positive pressure of 2 bar and at room temperature. This reservoir was connected to an evaporation device heated to 373 K via a liquid flow meter which allowed precise control of liquid methanol flows between 0.2 and 1.7 g hr$^{-1}$. Approximately 700 sccm (standard cubic centimetres per minute) of the Ar carrier gas flow was diverted into the evaporation device through a calibrated mass-flow controller, allowing the vaporized methanol to be carried into the reactor. The gas-phase methanol concentration was determined spectroscopically, by connecting the output of the CEM to a pressure controlled 10 cm long quartz cell at room temperature before entering the reactor. Methanol vapour concentrations were determined by absorption using the 185 nm line of a mercury pen-ray lamp and a solar blind channel photomultiplier tube (CPM) operating in photon counting mode. To ensure that only radiation from this specific mercury line was detected, an interference filter with a 10 nm FWHM transmission around 185 nm was placed in front of the CPM. To test for eventual fluctuations of the lamp and/or methanol vapour flows, the transmitted intensity was measured before, during and after each set of experiments at a specified methanol flow allowing the attenuated (I) and non-attenuated ($I_0$) intensities to be determined. The pressure within the cell was generally maintained at a value around 420 Torr, ensuring an adequate level of methanol absorption. The room temperature absorption cross-section of methanol was taken to be 6.05 × 10$^{-19}$ cm$^2$,[38] allowing its concentration to be calculated by the Beer-Lambert law. Methanol condensation between the absorption cell and the reactor was prevented by connecting the cell exit port to the reactor with a heated hose maintained at 353 K. Further condensation losses upstream of the Laval nozzle were neglected as the methanol vapour was diluted by at least a factor of five on entering this region, through mixing with the main Ar flow. CH$_3$OH concentrations as high as 5.0 × 10$^{14}$ cm$^{-3}$ were used during experiments at 296 K, while lower concentration ranges ($\leq$ 2.0 × 10$^{14}$ cm$^{-3}$) were used below room temperature.

For the CH$_3$CN experiments, acetonitrile vapour was introduced into the flow using a bubbler type method. Here, a small fraction of the Ar carrier gas was flowed into a bubbler containing CH$_3$CN at room temperature. At the exit of the bubbler, the CH$_3$CN vapour laden Ar was passed into a cold trap, held at 17 °C and a known pressure, before passing into the Laval nozzle reservoir through the same heated hose described above. It was not possible to determine the gas-phase CH$_3$CN concentration spectroscopically during these experiments, as its absorption cross-section is negligible at 185 nm. Despite this, the bubbler arrangement described above ensured that CH$_3$CN was maintained at its saturated vapour pressure value at 17 °C,[39] allowing the gas-phase CH$_3$CN concentration in the supersonic flow to be calculated precisely.

In common with our earlier studies of the reactions of ground state atomic carbon with CH$_3$OH[40] and CH$_3$CN,[41] the concentration ranges that could be used during these experiments were limited by the onset of cluster formation in the supersonic flow. Consequently, only those datapoints yielding a pseudo-first-order rate constant that displayed a linear dependence on the coreagent concentration were used in the final analysis to obtain temperature dependent rate constants. In particular, in the case of the O($^1$D) + CH$_3$CN reaction, it was not possible to perform a reliable measurement of the rate constant at 50 K due to the very small CH$_3$CN concentration range that could be employed, coupled with the large quenching contribution due to O($^1$D) quenching losses with the Ar carrier gas. For the same reason, it was not possible to perform H-atom product yield measurements for the O($^1$D) + CH$_3$CN reaction below room temperature, due to the low H-atom signals levels.

Ozone (O$_3$) photolysis, was used as the source of O($^1$D) atoms in the present work. Here, the 266 nm output of a pulsed (10 Hz) frequency quadrupled Nd:YAG laser (23 mJ/pulse) was directed along the central axis of the reactor, counterpropagating with the cold supersonic flow, producing a column of O($^1$D) atoms with uniform density along the entire length of the flow.

At this wavelength, O$_3$ photolysis yields mostly O($^1$D) atoms and a smaller amount of ground state O($^3$P) atoms,[42] although the O($^3$P) produced by photolysis (and by quenching of O($^1$D)) are unreactive with CH$_3$OH and CH$_3$CN at the temperatures of the present study. O$_3$ was generated by the UV irradiation of molecular oxygen in a quartz cell upstream of the Laval nozzle reservoir using a mercury lamp. O$_2$ molecules are photodissociated at these UV wavelengths, producing O($^3$P) atoms which undergo a termolecular association reaction with O$_2$ to form O$_3$. In order to increase the O$_3$ production yield, the cell pressure was maintained close to 760 Torr by a valve at the exit. The output of the O$_3$ generation cell was connected to an inlet port of the Laval nozzle reservoir while the valve and the connectors upstream of the O$_3$ cell were made from plastic to limit O$_3$ dissociation. O$_2$ concentrations in the range (3-7) × 10$^{13}$ cm$^{-3}$ were used during the present study. If we assume a 1% efficiency for O$_3$ production ([O$_3$] = (3-7) × 10$^{11}$ cm$^{-3}$) we expect O($^1$D) concentrations in the range (1-4) × 10$^{11}$ cm$^{-3}$ considering the photolysis energy, a beam diameter of 7 mm and an absorption cross-section for $O_3$ of $9.5 \times 10^{-18}$ $cm^2$.

The reaction kinetics were investigated during this work by following the disappearance of $O(^1D)$ atoms. Here $O(^1D)$ was detected directly by VUV laser induced fluorescence (VUV LIF) through its $2s^22p^4\ ^1D - 2s^22p^3(^2D°)3s\ ^1D°$ transition at 115.215 nm. For this purpose, a 10 Hz tunable UV laser at 345.65 nm (produced by frequency doubling the output of a Nd:YAG pumped dye laser around 691.3 nm) was focused into a cell containing Xe, allowing tunable VUV radiation to be generated by frequency tripling. In addition, Ar was added to the tripling cell to increase the VUV generation efficiency through phase matching. Several experiments were also performed to examine the product formation channels of the $O(^1D)$ + $CH_3OH$ and $O(^1D)$ + $CH_3CN$ reactions. Here, the $H(^2S)$ atom product of these reactions was detected at 121.567 nm in a similar manner to the $O(^1D)$ detection experiments except that Kr was used in the place of Xe. A $MgF_2$ lens was used to recollimate the VUV beam, while the residual UV beam remained divergent due to the difference in refractive index of $MgF_2$ at these two wavelengths.

In order to decrease the detection of scattered residual UV and VUV probe radiation by the detector, the tripling cell was not attached directly to the reactor. Instead, a 75 cm sidearm containing circular diaphragms was placed in between the two. In this way, most of the divergent UV radiation was blocked while the VUV probe beam was allowed to propagate into the reactor. Nevertheless, as this sidearm was open to the reactor it was always filled with residual gases. Attenuation of the VUV beam by absorption was reduced by continuously flushing the sidearm with Ar or $N_2$. In this configuration, the VUV probe beam crossed the cold supersonic flow at 90 degrees, exciting unreacted $O(^1D)$ atoms (or product $H(^2S)$ atoms). The resonant emission was collected by a LiF lens at right angles to both the flow and the probe laser, focusing it onto a solar blind photomultiplier tube (PMT). To avoid damage of the lens and the PMT window by reactive gases, both of these were isolated from the reactor by a LiF window. This region was evacuated to prevent further detection losses through atmospheric absorption of the VUV radiation. The output signal was integrated over time by a boxcar system (a gate of 20-40 ns was typically used) and averaged over 30 laser shots for each time point. The delay between the photolysis and probe lasers was controlled by a digital delay generator, which also served to synchronize the acquisition electronics. At least 100 different time points were recorded to establish each decay curve, with at least 15 time points recorded with the probe laser firing before the photolysis laser to determine the baseline level. The Ar carrier gas (99.999%) and $O_2$ precursor gas (99.999%) used to establish the supersonic flow during these experiments were regulated by calibrated mass-flow controllers, while flush gases $N_2$ (99.999%) and Ar were flowed into the sidearm through a needle valve. All gases were flowed directly from cylinders without further purification while $CH_3OH$ and $CH_3CN$ solvents were degassed prior to use.

## 3 Results and Discussion

**Rate Constants**

The photolysis of $O_3$ resulted in only low concentrations of $O(^1D)$ formed in the supersonic flow, while large excess concentrations of coreagents $CH_3OH$ and $CH_3CN$ with respect to $O(^1D)$ were employed during all experiments. As the loss of coreagent molecules through reaction with $O(^1D)$ was negligibly small, the pseudo-first-order approximation could be applied, with the $O(^1D)$ atoms decaying exponentially as a function of time. In Figure 1, some of the kinetic results of an experiment conducted at 127 K are shown. Here, the recorded $O(^1D)$ VUV LIF intensity is displayed as a function of time for experiments with (orange circles, $[CH_3OH]$ = $1.74 \times 10^{14}$ cm$^{-3}$) and without $CH_3OH$.

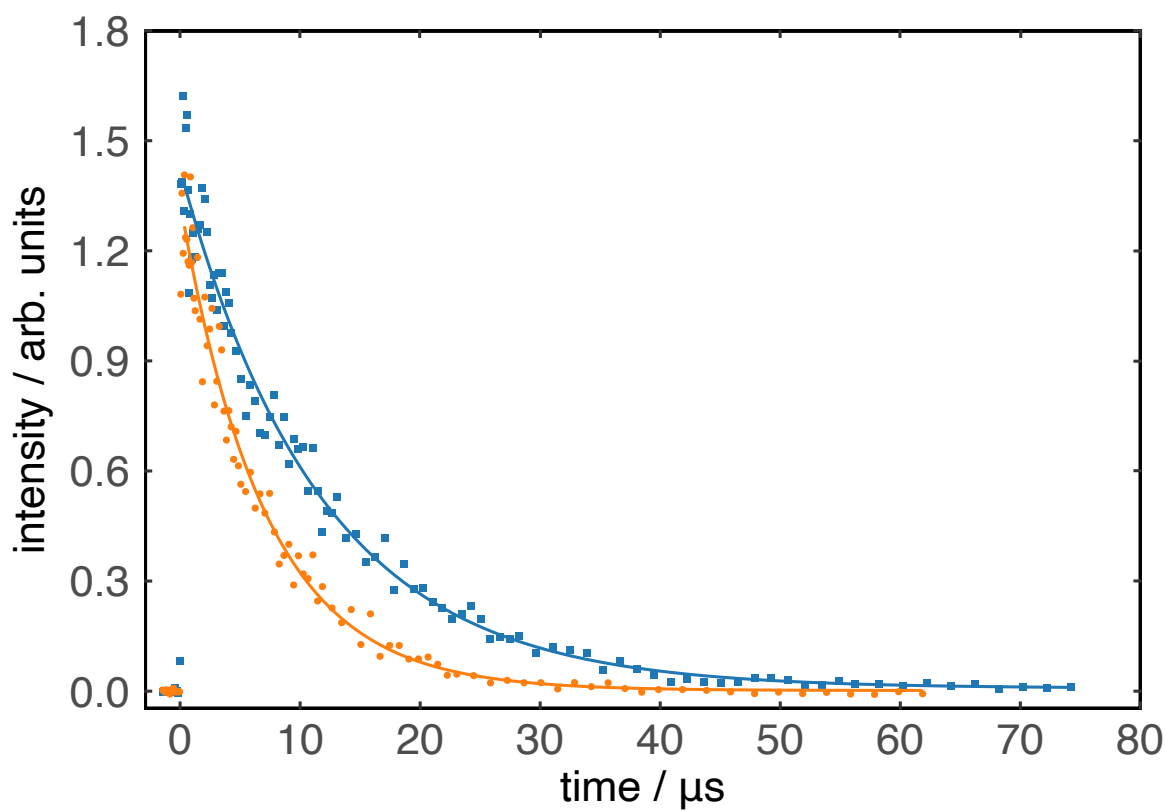

**Figure 1** O($^1$D) VUV LIF intensity as a function of time recorded at 127 K. (Orange circles) [CH$_3$OH] = 1.74 × 10$^{14}$ cm$^{-3}$ ; (blue squares) without CH$_3$OH. Solid lines represent exponential fits of the form $I(t) = I_0 \exp(-k_{1st}t)$ to the individual datasets (see text).

Similar curves were obtained for the temporal dependence of the O($^1$D) fluorescence signal in experiments performed with CH$_3$CN as the coreagent.

During these experiments, all decay profiles were analysed using a function of the form

$$I(t) = I_0 \exp(-k_{1st}t) \qquad (1)$$

where $I(t)$ and $I_0$ are the time dependent and initial LIF intensities respectively, $k_{1st}$ is the decay constant (the pseudo-first-order decay rate) and $t$ is the time. As the fluorescence signal from O($^1$D) atoms decays rapidly even in the absence of methanol, it is clear that several different processes contribute to $k_{1st}$. The most important of these is O($^1$D) removal by quenching collisions with the carrier gas Ar, which actually dominates O($^1$D) losses below room temperature, and reactive removal through collisions with the coreagent molecules. Other processes such as diffusion and removal by collisions with O$_2$ and O$_3$ molecules make only minor contributions to the overall loss rate of O($^1$D) atoms as determined by earlier work under similar conditions.[30] Decay curves such as those displayed in Figure 1 were recorded for a range of different coreagent concentrations at each temperature. The derived $k_{1st}$ values were plotted as a function of the coreagent concentration as shown in Figure 2 for CH$_3$OH and in Figure 3 for CH$_3$CN.

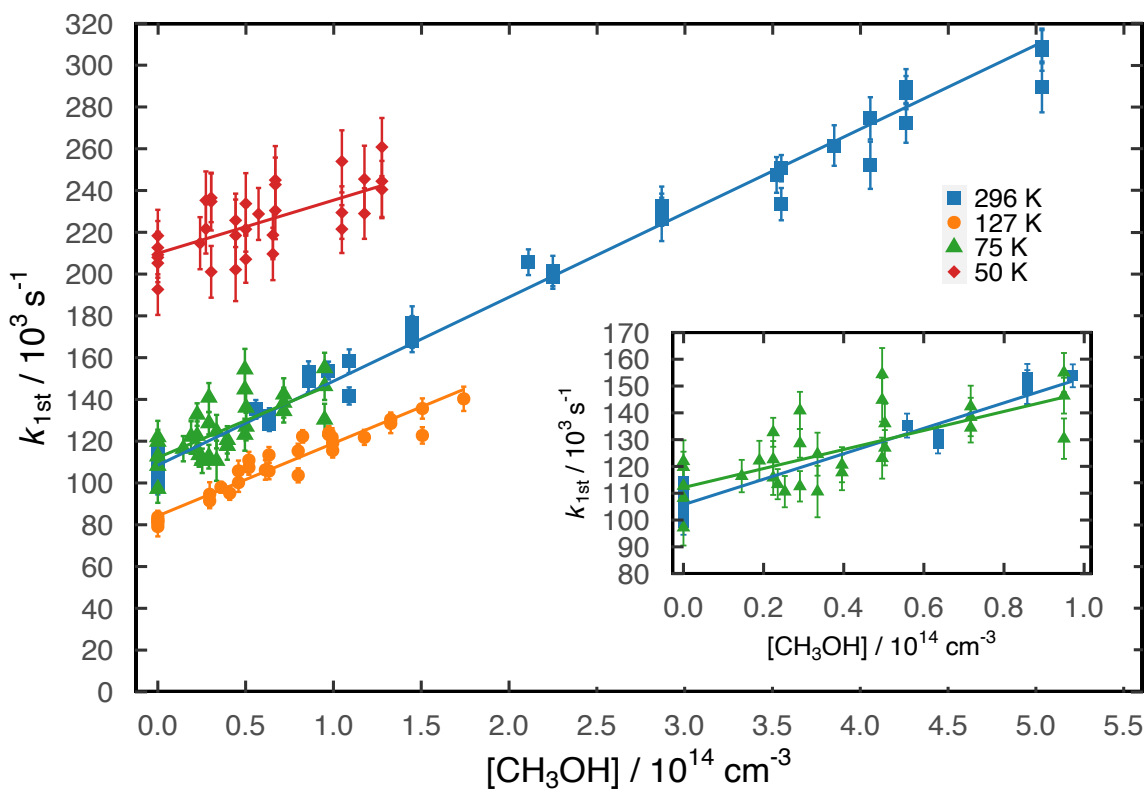

**Figure 2** Pseudo-first-order rate constants for O($^1$D) decay as a function of the methanol concentration. (Blue squares) data recorded at 296 K; (orange circles) data recorded at 127 K; (green triangles) data recorded at 75 K; (red diamonds) data recorded at 50 K. **Inset** - expanded view of the 75 K and 296 K data at low (< $1.0 \times 10^{14}$ cm$^{-3}$) [CH$_3$OH]. Weighted linear least squares fits to the individual data sets, as represented by solid lines, yield the second-order rate constant at a specified temperature. The error bars reflect the statistical uncertainty at the level of a single standard deviation obtained by exponential fits to decay profiles such as those shown in Figure 1.

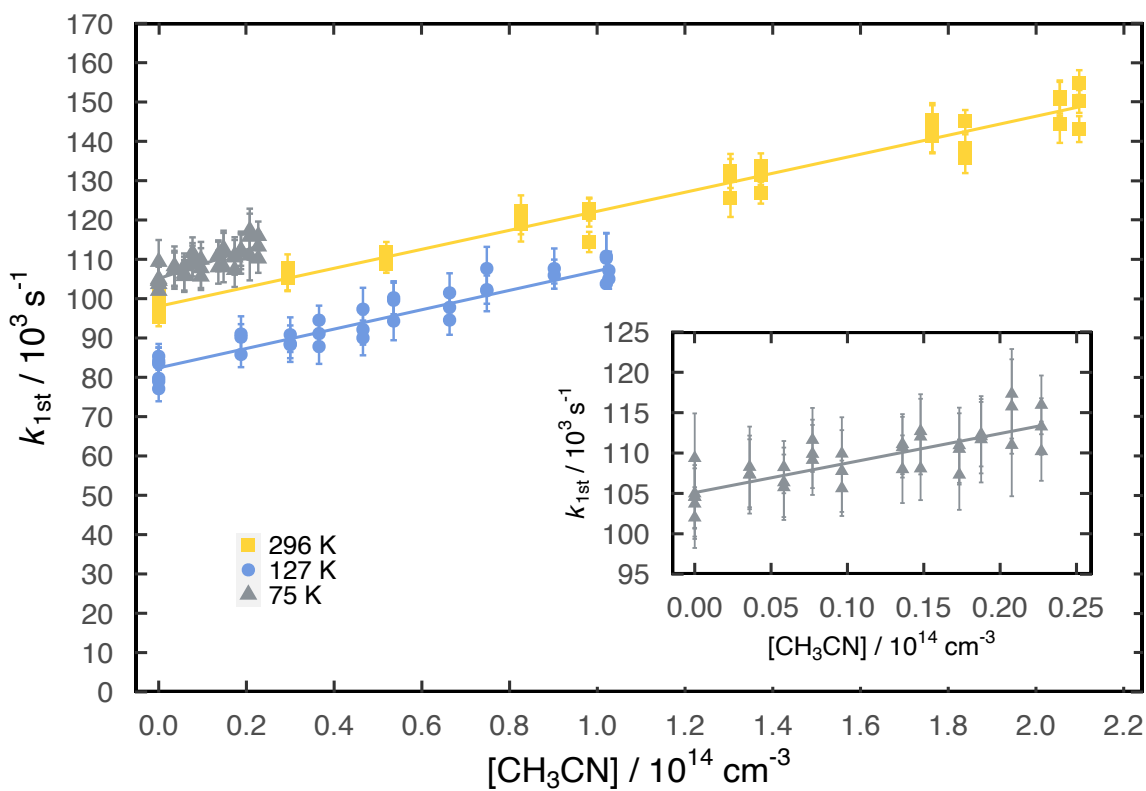

**Figure 3** Pseudo-first-order rate constants for O($^1$D) decay as a function of the acetonitrile concentration. (Yellow squares) data recorded at 296 K; (blue circles) data recorded at 127 K; (gray triangles) data recorded at 75 K. **Inset** - expanded view of the 75 K data. Weighted linear least-squares fits to the individual data sets, as represented by solid lines, yield the second-order rate constant at a specified temperature. The error bars reflect the statistical uncertainty at the level of a single standard deviation obtained by exponential fits to decay profiles such as those shown in Figure 1.

The solid lines shown in Figures 2 and 3 represent the weighted linear least-squares fits to the individual datasets. All fits display large y-axis intercept values, corresponding almost entirely to the quenching of O($^1$D) atoms by the carrier gas Ar, with negligibly small contributions from removal by $O_2$ and $O_3$. Nevertheless, as [Ar], [$O_2$] and [$O_3$] are all constant for any series of measurements, the changes in the derived pseudo-first-order rate constants arise due to the change in coreagent concentration alone. Consequently, these additional loss processes are not expected to modify the derived second-order rate constants for either of these reactions which are determined from the slopes of individual fits. In common with an earlier study on the C($^3$P) + $CH_3CN$ reaction,[41] cluster formation occurred readily in the experiments employing

CH$_3$CN due to its large electric dipole moment. Although it was possible to derive rate constants for the C($^3$P) + CH$_3$CN reaction to temperatures as low as 50 K, the large carrier gas quenching contribution of O($^1$D) atoms in the present work made measurements at this temperature unreliable. Considering a maximum usable CH$_3$CN concentration of 2.3 × 10$^{13}$ cm$^{-3}$ at 50 K, and a second-order rate constant that is likely to be in the range (3-4) × 10$^{-10}$ cm$^3$ s$^{-1}$ (based on the present measurements at higher temperature), the reactive contribution to $k_{1st}$ is expected to be in the range 7000-9000 s$^{-1}$. This compares to the expected quenching contribution by Ar of 1.8 × 10$^5$ s$^{-1}$ based on the measured quenching rate constant at 50 K[30] and a flow density of 2.6 × 10$^{17}$ cm$^{-3}$. Given the very small expected change in the derived $k_{1st}$ value, no attempt was made to study the C($^3$P) + CH$_3$CN reaction below 75 K.

The derived second-order rate constants for both these reactions are summarized in Table 1 with other relevant information and are displayed as a function of temperature in Figure 4, alongside previous work.

**Table 1** Measured second-order rate constants for the O($^1$D) + CH$_3$OH and O($^1$D) + CH$_3$CN reactions

| $T$ / K | [Ar] / 10$^{16}$ cm$^{-3}$ | $N^b$ | [CH$_3$OH] / 10$^{13}$ cm$^{-3}$ | $k_{O(^1D)+CH_3OH}$ / 10$^{-10}$ cm$^3$ s$^{-1}$ | $N^b$ | [CH$_3$CN] / 10$^{13}$ cm$^{-3}$ | $k_{O(^1D)+CH_3CN}$ / 10$^{-10}$ cm$^3$ s$^{-1}$ |
|---|---|---|---|---|---|---|---|
| 296 | 16.3 | 36 | 0 - 50.3 | (4.02 ± 0.43)$^c$ | 36 | 0 - 21.0 | (2.42 ± 0.27)$^c$ |
| 127 ± 2$^a$ | 12.6 | 30 | 0 - 17.4 | (3.48 ± 0.48) | 35 | 0 – 10.3 | (2.46 ± 0.36) |
| 75 ± 2 | 14.7 | 30 | 0 - 9.5 | (3.55 ± 1.10) | 36 | 0 - 2.3 | (3.66 ± 0.87) |
| 50 ± 1 | 25.9 | 33 | 0 - 12.8 | (2.54 ± 0.92) | | | |

$^a$Uncertainties on the temperatures represent the statistical (1σ) errors based on the fluctuations observed between different fixed positions during Pitot tube measurements of the impact pressure. These are the representative of the fluctuations $^b$Number of individual measurements. $^c$Uncertainties on the measured rate constants comprise the combined systematic errors (estimated to be 10%) and the derived statistical errors (1σ) obtained from the weighted fits shown in Figures 2 and 3. These 1σ values are multiplied by the appropriate t-distribution value corresponding to the 95 % confidence level to obtain the overall statistical

error. The statistical and systematic errors are combined using the expression $\sqrt{(\frac{u}{k})^2 + 0.1^2}$ where $k$ is the second-order rate constant and $u$ its associated statistical uncertainty.

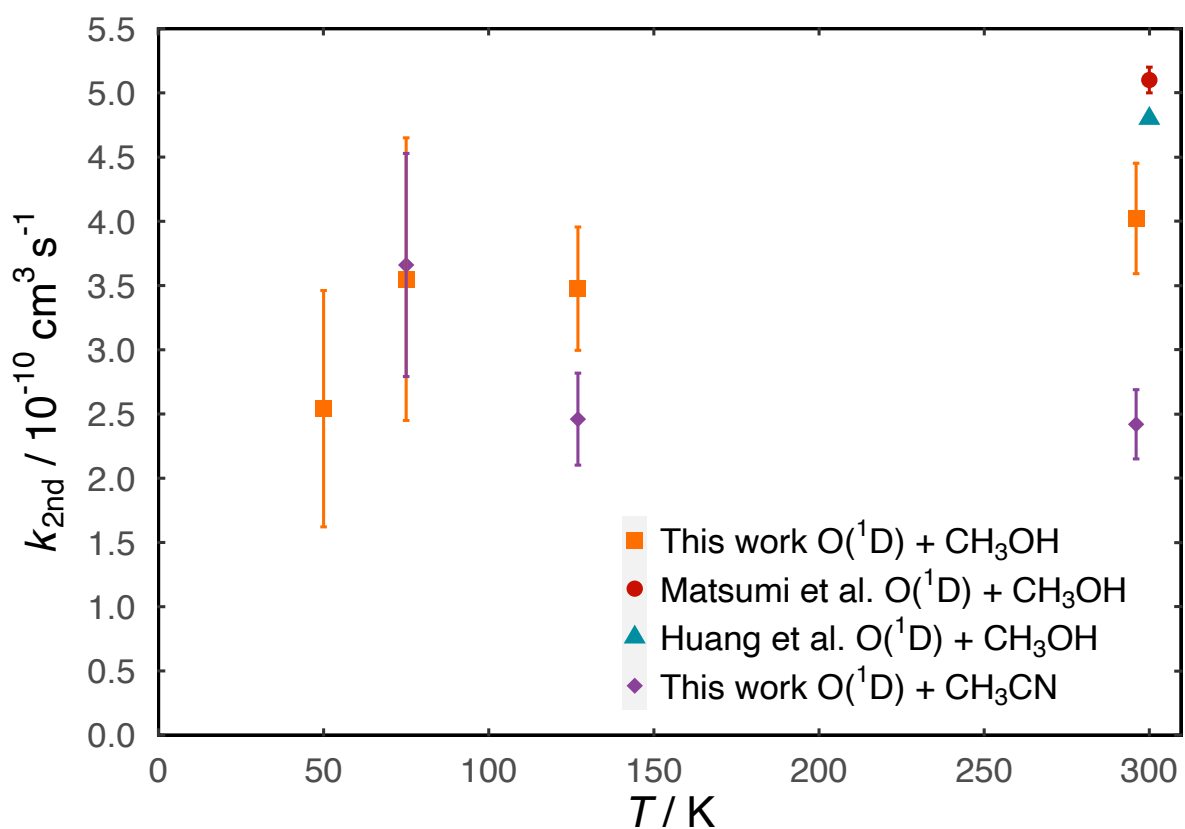

**Figure 4** Rate constants for the $O(^1D) + CH_3OH$ and $O(^1D) + CH_3CN$ reactions as a function of temperature. **The $O(^1D) + CH_3OH$ reaction:** (Red circle) Matsumi et al.[23]; (blue triangle) Huang et al.[24]; (orange squares) this work. **The $O(^1D) + CH_3CN$ reaction:** (purple diamonds) this work. The error bars on the present work represent the combined statistical uncertainty (derived as described above), with an added estimated systematic error of 10 %.

**Comparison with earlier work on the $O(^1D) + CH_3OH$ reaction**

In terms of earlier work, there are two previous kinetic studies of the $O(^1D) + CH_3OH$ reaction. Matsumi et al.[23] used the pulsed laser photolysis of both $N_2O$ at 193 nm and $O_3$ at 248 nm as the source of $O(^1D)$ atoms in their experiments on the kinetics and product channels of the $O(^1D) + CH_3OH$, $CH_3OD$ and $CD_3OH$ reactions. On the kinetics side, Matsumi et al.[23] measured a fast rate constant of $(5.1 \pm 0.1) \times 10^{-10}$ cm$^3$ s$^{-1}$ at 300 K for this process, following the loss of O($^1$D) atoms directly by LIF at 115.2 nm in the presence of an excess of CH$_3$OH. In an early relative rate measurement by Osif et al.,[43] these authors determined that the reaction was 5.5 ± 2 times faster than the O($^1$D) + N$_2$O reaction at room temperature, which would yield an absolute value for the rate constant greater than 6 × 10$^{-10}$ cm$^3$ s$^{-1}$. Huang et al.[24] used variational transition state theory to predict rate constants for the O($^1$D) + CH$_3$OH reaction based on new electronic structure calculations. Here, the geometries of the reagents, products, intermediates and transition states were optimized using density functional theory at the B3LYP/aug-cc-pVTZ level, while single point calculations were performed for these structures at the CCSD(T)/aug-cc-pVTZ level to obtain more reliable energies. They obtained a rate constant for total loss of O($^1$D) atoms of 4.8 × 10$^{-10}$ cm$^3$ s$^{-1}$, in excellent agreement with the measured value of Matsumi et al.[23] The room temperature rate constant measured during the present investigation of the O($^1$D) + CH$_3$OH reaction with a value of (4.02 ± 0.43) × 10$^{-10}$ cm$^3$ s$^{-1}$ is approximately 20 % smaller than the measured and calculated values of Matsumi et al.[23] and Huang et al.[24] respectively, although the rate constant is still very large; a clear indication of the barrierless nature of the process with reaction occurring at essentially every collision between reagents. At lower temperature, the rate constants measured during this study decrease slightly as the temperature falls reaching a value of (2.54 ± 0.92) × 10$^{-10}$ cm$^3$ s$^{-1}$ at 50 K. To the best of our knowledge, there are no earlier kinetic studies of this reaction below room temperature. An Arrhenius fit to the data yields the expression $k_{\mathrm{O}(^1\mathrm{D})+\mathrm{CH_3OH}}(T) = (4.43 \pm 0.02) \times 10^{-10} \exp(-25.06 \pm 7.60/T)$ cm$^3$ s$^{-1}$, valid over the 50-296 K range.

**Comparison with earlier work on the O($^1$D) + CH$_3$CN reaction**

There are no previous experimental studies of the O($^1$D) + CH$_3$CN reaction. On the theoretical side, Sun et al.[25] calculated several intermediates, transition states and product channels for the O($^1$D) + CH$_3$CN reaction during their study of the O($^3$P) + CH$_3$CN reaction using different levels of theory. They identified three intermediates at the G3(MP2)//B3LYP level of theory formed by the initial attack of the oxygen atom on the C-C and C-H bonds of CH$_3$CN followed by insertion, or by attack on the CN group to form an addition intermediate. As all these approaches of the oxygen atom were found to be barrierless, Sun et al.[25] suggested that the value of rate constant for the O($^1$D) + CH$_3$CN reaction could be comparable to those derived for other similar barrierless reactions such as O($^1$D) + CF$_3$CN[44] and O($^1$D) + CH$_3$Cl,[45] with room temperature rate constant values of (1.3 ± 0.2) × 10$^{-10}$ cm$^3$ s$^{-1}$ and (3.4 ± 0.1) × 10$^{-10}$ cm$^3$ s$^{-1}$ respectively. The rate constant values measured here with values in the range (2.4-3.7) × 10$^{-10}$ cm$^3$ s$^{-1}$ are certainly consistent with this idea and confirm the hypothesis that the O($^1$D) + CH$_3$CN reaction is also a barrierless process. At 127 K, the rate constant is essentially identical to the one obtained at 296 K, while at 75 K the rate constant increases to (3.66 ± 0.87) × 10$^{-10}$ cm$^3$ s$^{-1}$. Although the associated uncertainties suggest that this difference is real, with the rate constant value at 75 K being outside of the combined error bars of the three measured values, it may be that the overall experimental uncertainties has been underestimated for this temperature. Indeed, it can be seen from Figure 3 that the useful range of CH$_3$CN concentrations (due to CH$_3$CN cluster formation) is very small at 75 K compared to those used at higher temperatures. while the large y-intercept contribution due to O($^1$D) quenching by the carrier gas exacerbates the problem. Consequently, we recommend the use of a temperature independent value for the rate constant of this reaction of $k_{O(^1D)+CH_3CN}(T) = (2.85 \pm 0.70) \times 10^{-10}$ cm$^3$ s$^{-1}$.

**H-atom Product Yields**

In previous work on the reactions of C($^1$D) and C($^3$P) atoms with hydrogen bearing molecules, [46,41] absolute H-atom product branching ratios were obtained by comparing the H-atom formation curves produced by the target reaction to the ones produced by a reference reaction with a known H-atom yield. During these experiments, the excess coreagent concentration was varied so that it was possible to compare traces with similar H-atom production rates. Under these conditions, the non-reactive losses (essentially diffusional losses) of the minor coreagent (C($^1$D) or C($^3$P)) and the H-atom product were identical for each pair of curves so that absolute H-atom yields could be derived directly from the observed intensities. In the case of O($^1$D) atom reactions, the O($^1$D) + Ar → O($^3$P) + Ar quenching reaction is faster,[30] so that if the same procedure as used for the C($^1$D) and C($^3$P) atom reactions was used here, any minor error in matching the H-atom formation rates for the target and reference reactions could result in large discrepancies in the absolute H-atom yields. In order to prevent this issue, an alternative method as described by Nuñez-Reyes et al.[47,48] has been applied in other recent work[49] and here. In the present study, the atomic hydrogen yields from the target O($^1$D) + CH$_3$OH and O($^1$D) + CH$_3$CN reactions were measured relative to the yield of the O($^1$D) + H$_2$ reference reaction; a process that is assumed to produce 100 % H-atoms. During these experiments, the peak H-atom VUV LIF signal, $I_{Hmax}$, is recorded for several concentration values of the target reaction coreagent (CH$_3$OH or CH$_3$CN) and the reference reaction coreagent (H$_2$). The times (the delay between photolysis and probe lasers) corresponding to the appearance of the peak intensity values were calculated prior to the experiments using the expression

$$t_{max} = \frac{1}{k_{L(H)} - k'} \ln \left(\frac{k_{L(H)}}{k'}\right) \qquad (2)$$

where $k_{L(H)}$ is the diffusional loss rate constant for H-atoms derived from earlier work under similar conditions and $k'$ is the pseudo-first-order rate constant for H-atom formation (equal to the pseudo-first-order rate constant for O($^1$D) loss as measured during the kinetic experiments, with $k' = k_{O(^1D)+X}[X] + k_{O(^1D)+Ar}[Ar]$. Here, X = CH$_3$OH, CH$_3$CN or H$_2$ and $k_{O(^1D)+X}$ and $k_{O(^1D)+Ar}$ are the rate constants for reaction with X and quenching by Ar respectively. The expression for $I_{Hmax}$ can be written

$$I_{Hmax} = \left(\frac{k'_H [O(^1D)]_0}{k'}\right) \times (1 - k_{L(H)} t_{max}) \qquad (3)$$

where $[O(^1D)]_0$ is the initial concentration of O($^1$D) atoms and $k'_H$ is the pseudo-first-order rate constant for the H-atom production channel(s). In the case of H$_2$, $k_{O(^1D)+X}[H_2] = k'_H$ as the reaction is assumed to yield 100 % H-atoms.

The $I_{Hmax}$ values were then corrected for absorption of the VUV probe laser and VUV emission by the coreagent molecules (less than 6 % difference for both reactions). For the H-atom yield experiments on the O($^1$D) + CH$_3$OH reaction, an additional correction was applied to account for the formation of H-atoms by the photodissociation of a small fraction of CH$_3$OH at 266 nm. This magnitude of this correction was derived from separate experiments performed at the same delay times and with the same CH$_3$OH concentrations as those used in the H-atom yield experiments, but in the absence of O($^1$D) atoms. This correction accounted for as much as 44 % of the total H-atom signal in the least favorable instance (room temperature experiments employing [CH$_3$OH] = 4.3 × 10$^{14}$ cm$^{-3}$). A typical plot of the corrected and uncorrected $I_{Hmax}$ values versus time for the O($^1$D) + CH$_3$OH reaction measured at 296 K is shown alongside the corresponding $I_{Hmax}$ values for the O($^1$D) + H$_2$ reference reaction in Figure 5.

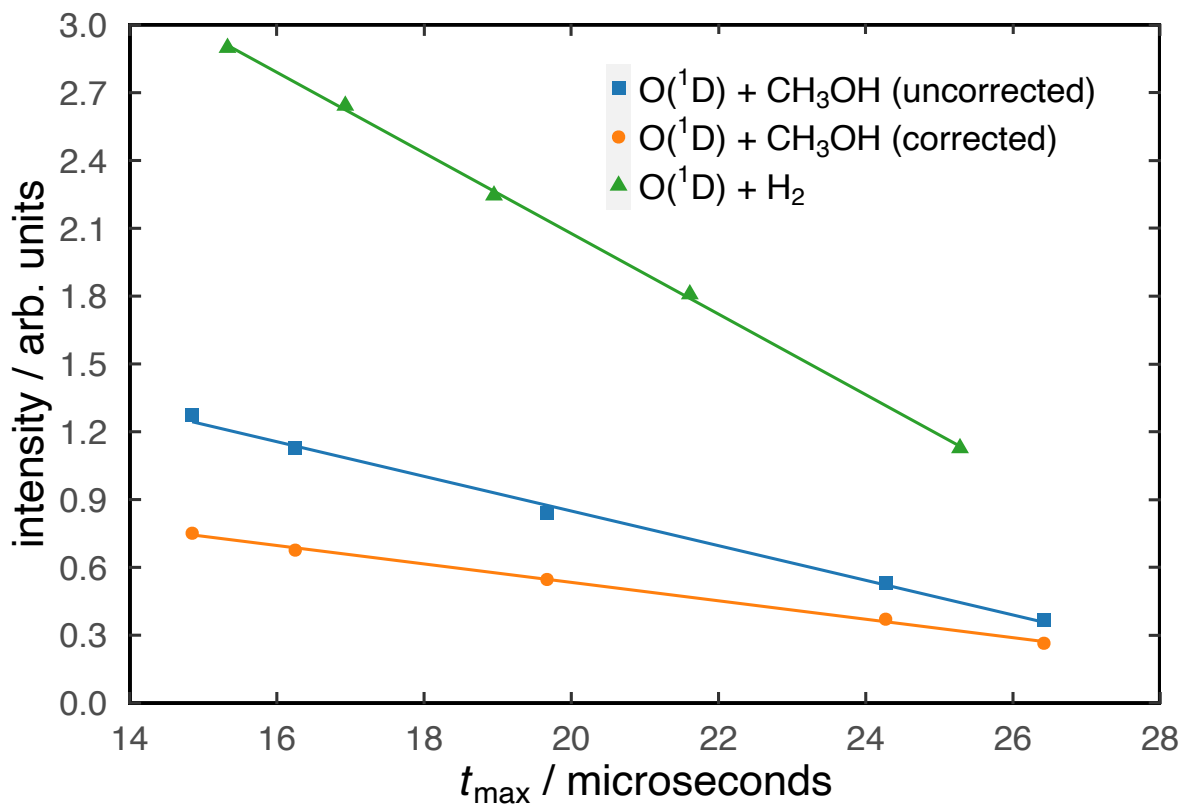

**Figure 5** Maximum H-atom product intensity, $I_{Hmax}$, as a function of maximum time, $t_{max}$, recorded at 296 K, with each point corresponding to a different methanol concentration. (Blue squares) O($^1$D) + CH$_3$OH reaction; (orange circles) O($^1$D) + CH$_3$OH reaction corrected for H-atom production by CH$_3$OH photolysis and absorption losses by CH$_3$OH; (Green triangles) O($^1$D) + H$_2$ reaction.

After subtraction of the photolysis contribution to the H-atom signal and correction for absorption of the VUV probe laser and fluorescence emission by CH$_3$OH, the absolute H-atom yield is determined from the ratio of the slopes of the two straight line plots. It should be noted at this point that the due to the O($^1$D) quenching contribution $k_{O(^1D)+Ar}[Ar]$ to the overall rate constant for O($^1$D) loss (or H-atom formation) for both the target and reference reactions, care should be taken when applying this method as numerical simulations indicate that the analysis procedure only yields accurate H-atom yields over a limited range of concentrations. Nevertheless, as all of the parameters needed to simulate $I_{Hmax}$ and $t_{max}$ are known (knowledge of $[O(^1D)]_0$ is not required as this parameter is identical for the target and reference reactions and cancels out in the analysis), it is possible to use the simulations to correct the derived yields if required.

An average absolute H-atom yield of 0.205 ± 0.06 is obtained at 296 K while at 127 K, a similar analysis gives an average absolute H-atom yield of 0.200 ± 0.03 directly from the ratio of the slopes. Our simulations indicate that the final H-atom yields should be 0.204 ± 0.06 and 0.198 ± 0.03 at 296 K and 127 K respectively after the numerical correction is applied.

In the case of the O($^1$D) + CH$_3$CN reaction, H-atom yield measurements were only performed at 296 K. Here, no correction was required to account for CH$_3$CN photolysis at 266 nm, due to its negligibly small absorption cross section at this wavelength, although a maximum correction of 5 % was applied to account for VUV absorption by CH$_3$CN. These experiments determined an average absolute H-atom yield of 0.183 ± 0.03 which is corrected to 0.177 ± 0.03. The corrected product branching ratios are summarized in Table 2.

**Table 2** H-atom yields for the O($^1$D) + CH$_3$OH and O($^1$D) + CH$_3$CN reactions

| T / K | N[a] | H-atom yield O($^1$D) + CH$_3$OH | N[a] | H-atom yield O($^1$D) + CH$_3$CN |
|---|---|---|---|---|
| 296 | 4 | 0.204 ± 0.06[b] | 4 | 0.177 ± 0.03 |
| 127±2[c] | 8 | 0.198 ± 0.03 | | |

[a]Number of branching ratios determinations. [b]The error bars reflect the statistical uncertainties at the 95% confidence level. [a]Uncertainties on the temperatures represent the statistical (1σ) errors based on the fluctuations observed between different fixed positions during Pitot tube measurements of the impact pressure.

**The O($^1$D) +CH$_3$OH reaction**

To date, product yield studies have only been performed at room temperature for the O($^1$D) + CH$_3$OH reaction and its isotopologues CD$_3$OH, CH$_3$OD and CD$_3$OD. Goldstein & Wiesenfeld[22] derived a yield of 0.58 ± 0.02 for OH formation for the O($^1$D) + CH$_3$OH reaction (relative to the O($^1$D) + H$_2$O reaction) by following OH radical production by LIF over the 306-319 nm range. These authors used pulsed laser photolysis of O$_3$ at 266 nm as the source of O($^1$D) atoms in their experiments in a similar manner to the present work. They also determined total yields (OH + OD) for the three other isotopologues allowing them to conclude that the primary site for O($^1$D) attack was the O-H (O-D) bond rather than one of the C-H bonds. In addition to their kinetic studies of the O($^1$D) + CH$_3$OH reaction described earlier, Matsumi et al.[23] also investigated the product channels of the O($^1$D) + CH$_3$OH, CD$_3$OH and CH$_3$OD reactions at room temperature. By using a 1:1 mixture of CH$_3$OH and D$_2$ in the presence of O($^1$D) atoms they were able to measure a [H]/[D] isotopic signal ratio = 0.97 ± 0.1 by following the formation of H-(D-) atoms by pulsed LIF at the Lyman-α wavelength. The quantum yield for H-atom production by the O($^1$D) + CH$_3$OH reaction was then estimated from the product of the isotopic signal ratio and the ratio of the rate constants for the O($^1$D) + CH$_3$OH and O($^1$D) + D$_2$ reactions, with a value equal to 0.18 ± 0.02. Based on their experimental H- (D-) atom yields from experiments employing CD$_3$OH, Matsumi et al.[23] observed that the production of D-atoms was four times greater than the production of H-atoms, suggesting that the intermediate diol species formed, DO-CD$_2$-OH, preferentially falls apart at the new O-D bond rather than at the original O-H bond. Nevertheless, these authors could not exclude the possibility that H-(D-) atom production could arise from a combination of both C-H and O-H bond cleavage.

In addition to their theoretical kinetics study of the O($^1$D) + CH$_3$OH reaction, Huang et al.[24] also performed an experimental investigation of OH and OD formation from the reaction of O($^1$D) atoms with both CD$_3$OH and CH$_3$OD through time-resolved Fourier-transform IR emission spectroscopy. They concluded that the H-atom abstraction channels were unimportant (based on their theoretical work) while the extensive vibrational excitation they observed in OH produced from O($^1$D) insertion into a C-H bond of the methyl group in CH$_3$OD indicated that the intermediate OH*-CH$_2$OD lifetime is too short for complete intramolecular vibrational relaxation to occur, so that the reaction energy is effectively localized near the newly formed bond, in good agreement with the conclusions of Matsumi et al.[23] This conclusion was further supported by the observation that OD produced by fission of the old C-OD bond was less vibrationally excited. These authors hypothesized that O($^1$D) attack occurs preferentially at the methyl site rather than at the hydroxyl site, despite the fact that the observed [OH]/[OD] ratio indicated a preference for OH formation from attack at the hydroxyl group. They argued that this discrepancy might be attributed to the fact that the energized HOCH$_2$OH* intermediate formed by attack at the methyl group could also dissociate to form atomic hydrogen. The H-atom yield measured at 296 K in the present study of 0.204 ± 0.06 is in excellent agreement with the value derived by Matsumi et al.,[23] and supports the conclusions of both Matsumi et al.[23] and Huang et al.[24] As the H-atom yield derived here does not seem to vary significantly as a function of temperature with a measured H-atom yield of 0.198 ± 0.03 at 127 K, this result could be another indication of the preferential production of atomic hydrogen from O($^1$D) insertion into one of the C-H bonds of the methyl group. Indeed, Huang et al.[24] calculated the energies of the possible channels leading to H elimination at the CCSD(T)//B3LYP/aug-cc-pVTZ level, finding that H-atom production from O($^1$D) attack at the methyl group was a highly exothermic process (these channels, leading to OCH$_2$OH + H and HOCHOH + H were found to be exothermic by 204.6 kJ/mol and 237.7 kJ/mol respectively), whereas H-atom production from O($^1$D) attack at the hydroxyl site was found to be only slightly exothermic (-23.4 kJ/mol) which would make this channel less favourable at low temperature.

**The O($^1$D) +CH$_3$CN reaction**

According to the theoretical study by Sun et al.,[25] after formation of the insertion (HOCH$_2$CN, CH$_3$OCN) or addition (CH$_3$C(O)N) intermediates following O($^1$D) attack, these species subsequently evolved either directly or through other transitions states and intermediates over the singlet surface to form products. The major products were predicted to be CH$_3$ + NCO and CH$_2$CN + OH with H$_2$O + HCCN also identified as possible minor products. Interestingly, no discussion was made of the possibility for the formation of atomic hydrogen (with HOCHCN, OCH$_2$CN and CH$_2$NCO as coproducts) as potential product channels over the singlet surface. Despite this, earlier work by Matsumi et al.[45] on the closely related O($^1$D) + CH$_3$Cl reaction clearly demonstrated that low yields of H atoms (around 7 %) were produced for this process, suggesting that this might also be a possible exit channel for the O($^1$D) + CH$_3$CN reaction. In order to clarify the origin of the product hydrogen atoms observed in the present experiments, we ran supplementary electronic structure calculations to elucidate the possible H-atom formation pathways over the singlet potential energy surface.

For these bond dissociation calculations, we first used Complete Active Space Self-Consistent Field (CASSCF) calculations using 6 active orbitals and 6 active electrons. With the resulting molecular orbitals, the Davidson corrected Multi-Reference Configuration Interaction energies (MRCI+Q) with the same active space were calculated using the MOLPRO suite of programs[50] employing an augmented double zeta atomic basis set, aug-cc-pVDZ (AVDZ). During the calculations all distances (except the dissociated bond) and angles were optimized at the CASSCF level. Three possible pathways for hydrogen atom production have been found, although it is possible that others also exist. The first two result in the production of H + HOCHCN and H + OCH$_2$CN products from the HOCH$_2$CN adduct (denoted s-IM8 by Sun et al.[25] and formed by O($^1$D) addition to one of the C-H bonds of CH$_3$CN). The third pathway is the production of H + CH$_2$NCO from the CH$_3$NCO adduct (denoted s-IM2 by Sun et al.[25] and formed by rearrangement from CH$_3$C(O)N, denoted s-IM1, which is formed by O($^1$D) addition to the C-atom of the cyano group). The MRCI+Q/AVDZ calculations clearly show that none of these pathways present barriers during C-H or O-H bond dissociation. Indeed, the absence of a barrier for these processes is characteristic of homolytic bond fission on singlet surfaces leading to two radicals in doublet ground states. The results of the MRCI+Q/AVDZ calculations for these three bond dissociation pathways can be found in the supplementary information file.

In addition to the MRCI calculations of the bond dissociation pathways, density functional theory (DFT) calculations with the M06-2X functional[51] at the aug-cc-pVTZ (AVTZ) level also show that all three dissociation channels are strongly exothermic.

O($^1$D) + CH$_3$CN → HOCH$_2$CN → H + HOCHCN        ΔH$_r$ = -206 (-152) kJ/mol        (3)

→ HOCH$_2$CN → H + OCH$_2$CN        ΔH$_r$ = -164 (-81) kJ/mol        (4)

→ CH3C(O)N → CH$_3$NCO → H + CH$_2$NCO        ΔH$_r$ = -216 (-147) kJ/mol        (5)

Here, the reported energies are relative to O($^3$P) + CH$_3$CN, with the experimental value of the O($^1$D)-O($^3$P) energy difference added on, as DFT calculations using the M06-2X functional at the AVTZ level do not give accurate results for the energy of atomic oxygen in its $^1$D state, which is not well described by a single-determinant wavefunction as implemented in hybrid meta-GGA functionals such as M06-2X. The energy values reported in brackets are those derived in the MRCI+Q/AVDZ calculations.

The three major pathways identified by Sun et al.,[25] over the singlet surface leading to either CH$_2$CN + OH or CH$_3$ + NCO as products, were found to be exothermic by 229 and 254 kJ/mol respectively at the G3(MP2)//B3LYP level of theory, with all three pathways involving a single intermediate species (s-IM8, s-IM1 or CH$_3$OCN, denoted as s-IM5) and no TS structures between the intermediate species and the products. This can be compared with the pathways found during the present investigation, with channel (3) presenting similar characteristics to the major pathways characterized by Sun et al.[25] In contrast, channel (4) is significantly less exothermic than channel (3) and those described by Sun et al.[25] and might therefore be less important than these pathways. Channel (5) involves two intermediate species, and although the products H + $CH_2NCO$ are similar in energy to channel (3) it might be expected that this pathway is less kinetically favourable because it involves an additional TS structure (s-TS1). Considering the various possible pathways for the $O(^1D)$ + $CH_3CN$ reaction, the major products are likely to be $CH_3$ + NCO, followed by $CH_2CN$ + OH, H + HOCHCN and $H_2O$ + HCCN as other potentially important channels, although this hypothesis would need to be confirmed through transition state theory calculations of the unimolecular dissociation kinetics of the various intermediate species and/or through new experimental studies. Despite this, the measured H-atom yield close to 0.2 obtained in the present work appears to support this hypothesis.

## 4 Conclusions

This paper presents the results of an experimental kinetic study of the reactions between atomic oxygen in its first excited state and the coreagents methanol and acetonitrile. A continuous supersonic flow reactor was used to perform the measurements, allowing these processes to be investigated at a range of temperatures between 50 and 296 K. $O(^1D)$ was generated by pulsed laser photolysis of $O_3$ at 266 nm, while pulsed laser induced fluorescence was used to detect these atoms at 115.2 nm. The measured rate constants were seen to be large for both reactions with values greater than $2 \times 10^{-10}$ $cm^3$ $s^{-1}$ at all temperatures. The $O(^1D)$ + $CH_3OH$ and $O(^1D)$ + $CH_3CN$ reactions are seen to display contrasting temperature dependences however. While rate constants for the former process were seen to increase with temperature, the measured rate constants for the latter process are mostly independent of temperature considering the overall experimental uncertainties. In addition to the kinetic study, the product channels were also investigated by comparing H-atom yields for both reactions with the H-atom yield of the $O(^1D)$ + $H_2$ reference reaction. Although the derived H-atom product yields for the $O(^1D)$ + $CH_3OH$ reaction can be adequately explained through previous experimental and theoretical work, there are no earlier experimental or theoretical studies of H-atom formation by the $O(^1D)$ + $CH_3CN$ reaction. Consequently, new electronic structure calculations have been performed to elucidate the possible H-atom formation channels, which are found to be exothermic and barrierless. These pathways are discussed in the context of previous work examining several alternative product channels for this reaction.

**Author Information**

**Corresponding Author**

*Email: kevin.hickson@u-bordeaux.fr.

**Acknowledgements**

K. M. H. acknowledges support from the French program ''Physique et Chimie du Milieu Interstellaire'' (PCMI) of the CNRS/INSU with the INC/INP co-funded by the CEA and CNES as well as funding from the ''Programme National de Planétologie'' (PNP) of the CNRS/INSU.